\documentclass[12pt,a4paper]{article}
\usepackage{amsfonts,latexsym}
\usepackage{amsmath,amssymb}
\usepackage{graphicx,color}

\renewcommand{\thefootnote}{\fnsymbol{footnote}}

\begin{document}

\vspace{12mm}

\begin{center}
{{{\Large {\bf Scalar hairy black holes  \\ in Einstein-Maxwell-conformally coupled scalar theory }}}}\\[10mm]

{De-Cheng Zou$^{a,b}$\footnote{e-mail address: dczou@yzu.edu.cn} and Yun Soo Myung$^a$\footnote{e-mail address: ysmyung@inje.ac.kr}}\\[8mm]

{${}^a$Institute of Basic Sciences and Department  of Computer Simulation, Inje University Gimhae 50834, Korea\\[0pt] }

{${}^b$Center for Gravitation and Cosmology and College of Physical Science and Technology, Yangzhou University, Yangzhou 225009, China\\[0pt]}
\end{center}
\vspace{2mm}

\begin{abstract}
We obtain scalar hairy black holes from Einstein-Maxwell-conformally coupled scalar (EMCS) theory with the scalar coupling parameter $\alpha$ to the Maxwell term.
In case of $\alpha=0$, the $\alpha=0$ EMCS  theory provides constant (charged) scalar hairy black hole and charged BBMB (Bocharova-Bronnikov-Melnikov-Bekenstein) black hole
where the former is stable against full perturbations, while the latter remains unstable because it belongs to an extremal black hole.
It is noted that for $\alpha\not=0$, the unstable Reissner-Nordstr\"{o}m black holes without scalar hair imply infinite branches  of $n=0(\alpha \ge 8.019),1(\alpha \ge 40.84),2(\alpha \ge 99.89),\cdots$ scalarized charged black holes.
In addition, for $\alpha>0$, we develop a single branch of  scalarized charged black hole solutions  inspired by  the constant scalar hairy black hole.
Finally, we obtain the numerical charged BBMB black hole solution from the $\alpha=0$ EMCS theory.
\end{abstract}
\vspace{5mm}

\vspace{1.5cm}

\hspace{11.5cm}{Typeset Using \LaTeX}
\newpage
\renewcommand{\thefootnote}{\arabic{footnote}}
\setcounter{footnote}{0}


\section{Introduction}
No-hair theorem implies that a black hole is completely described by the mass $M$, electric charge $Q$, and angular momentum $J$~\cite{Ruffini:1971bza}.
Outside the horizon, Maxwell and gravitational fields satisfy the Gauss-law.

A minimally coupled scalar does not obey the Gauss-law and thus, a black hole
cannot have a scalar hair~\cite{Herdeiro:2015waa}. The scalar-tensor theories admitting other than the Einstein gravity are rather rare. However, considering  the Einstein-conformally coupled scalar  theory leads to a secondary scalar hair around the BBMB black hole
although the scalar hair blows up on the horizon~\cite{Bocharova:1970skc,Bekenstein:1974sf}. In this case, the scalar equation becomes $\nabla^2 \phi=0$ for a minimally coupled scalar when imposing the condition of $R=0$.
 This was considered as the first counterexample to the no-hair conjecture for black holes. On 1991,  Xanthopoulos and Zannias have pointed out that the BBMB black hole is the unique static and  asymptotically flat solution to the Einstein-conformally coupled scalar theory~\cite{Xanthopoulos:1991mx}.
 It  may remain an open problem that could be resolved numerically determining the exact nature of the BBMB solution.
 Very recently, adding the Weyl squared term to this action led to  a non-BBMB black hole solution with a primary scalar hair where we found the BBMB black hole solution numerically  when turning off the Weyl squared term~\cite{Myung:2019adj}.

On the other hand, scalarized (charged) black holes were found  from the Einstein-Gauss-Bonnet-Scalar (EGBS) theory~\cite{Doneva:2017bvd,Silva:2017uqg,Antoniou:2017acq} (Einstein-Maxwell-Scalar (EMS) theory~\cite{Herdeiro:2018wub})
by introducing  the quadratic and exponential couplings of a  scalar to the Gauss-Bonnet term $f(\phi){\cal G}$ (Maxwell term $f(\phi) F^2$). In these models, we mention that the scalar is non-minimally coupled to the system. In this approach of spontaneous scalarization, the linearized scalar equation  is important to determine infinite branches of the $n=0,1,2,\cdots$ scalarized (charged) black holes.

In this work, we wish to introduce a combined model of  Einstein-Maxwell-conformally coupled scalar (EMCS) theory to obtain various scalarized charged black hole solutions.
In case of $\alpha=0$, the EMCS theory  without scalar coupling (namely, $\alpha=0$ EMCS theory) provides  the constant scalar hairy  black hole and charged BBMB (Bocharova-Bronnikov-Melnikov-Bekenstein) black hole.
We show that  the former is stable against full perturbations. We mention that  the latter remains unstable because it belongs to an extremal black hole~\cite{Bronnikov:1978mx}.
For $\alpha\not=0$, it is interesting to note that  the unstable Reissner-Nordstr\"{o}m (RN) black holes without scalar hair imply the appearance of $n=0(\alpha \ge 8.019),1(\alpha \ge 40.84),2(\alpha \ge 99.89),\cdots$ scalarized charged black holes, as was shown  in the EMS theory~\cite{Myung:2018vug}.
Importantly, for $\alpha>0$, we obtain  a single branch of scalarized charged black holes inspired by  the constant scalar hairy black hole.
Finally, we show that  the charged BBMB black hole can be found  numerically  in the $\alpha=0$ EMCS theory when imposing the asymptotically flat condition.

\section{EMCS theory}
We start with the Einstein-Maxwell-conformally coupled scalar (EMCS) theory  whose action is given by
\begin{eqnarray}S_{\rm EMCS}=\frac{1}{16 \pi G}\int d^4 x\sqrt{-g}
\Big[R-f(\phi)F_{\mu\nu}F^{\mu\nu}-\beta\Big(\phi^2R+
6\partial_\mu\phi\partial^\mu\phi\Big)\Big],
\label{EMCS}
\end{eqnarray}
where $f(\phi)=1+\alpha \phi^2$  includes `$\alpha \phi^2$' (quadratic  scalar coupling term with coupling parameter $\alpha$) and  the last term corresponds to the conformally coupled scalar action with parameter $\beta$. In the limit of $\alpha\to 0$, the above action recovers  the $\alpha=0$ EMCS theory.
It was found by Bekenstein~\cite{Bekenstein:1974sf} that a solution $(\hat{g}_{\mu\nu},\hat{A}_\mu,\hat{\phi})$ of Einstein-Maxwell-minimally coupled scalar theory
could be mapped to a solution $(g_{\mu\nu},A_\mu,\phi)$ to  the $\alpha=0$ EMCS theory by the conformal transformations: $\hat{\phi}\to \phi=\sqrt{1/\beta}\tanh(\sqrt{\beta} \hat{\phi})$, $\hat{A}_\mu \to A_\mu=\hat{A}_\mu$, $\hat{g}_{\mu\nu}\to g_{\mu\nu}=(1-\sqrt{\beta}\phi^2)^{-1}\hat{g}_{\mu\nu}$. Using this transformation,
Bekenstein has  discovered the charged BBMB black hole~\cite{Astorino:2013xc}. Also, Astorino~\cite{Astorino:2013sfa} has recently found  the constant scalar hairy black hole solution.
In this work,  we choose $\beta=1/3$ and $G=1$ for simplicity.

We derive the Einstein equation from (\ref{EMCS})
\begin{equation} \label{nequa1}
G_{\mu\nu}=2(1+\alpha \phi^2) T^{\rm M}_{\mu\nu}+T^{\rm \phi}_{\mu\nu},
\end{equation}
where the Einstein tensor  is given by $G_{\mu\nu}=R_{\mu\nu}-Rg_{\mu\nu}/2$.
The energy-momentum tensors for Maxwell theory and  conformally coupled scalar theory  are  defined by
\begin{eqnarray} \label{equa2}
T^{\rm M}_{\mu\nu}&=&F_{\mu\rho}F_{\nu}~^\rho- \frac{F^2}{4}g_{\mu\nu},\label{trace} \\
T^{\rm \phi}_{\mu\nu}&=&\beta\Big[\phi^2G_{\mu\nu}+g_{\mu\nu}\nabla^2(\phi^2)-\nabla_\mu\nabla_\nu(\phi^2)+6\nabla_\mu\phi\nabla_\nu\phi-3(\nabla\phi)^2g_{\mu\nu}\Big],\nonumber
\end{eqnarray}
where we observe the traceless condition of $T^{{\rm M} \mu}_\mu=0$. The Maxwell equation takes the form
\begin{equation} \label{maxwell-eq}
\nabla^\mu F_{\mu\nu}=2\alpha \phi \nabla^{\mu}(\phi) F^2.
\end{equation}
Finally, the scalar
equation is given by
\begin{equation} \label{ascalar-eq}
\nabla^2\phi-\frac{1}{6}R\phi-\frac{\alpha}{6\beta}  F^2 \phi=0.
\end{equation}
Taking the trace of the Einstein equation (\ref{nequa1}) together with (\ref{ascalar-eq}) leads to a non-vanishing Ricci scalar as
\begin{equation} \label{ricciz}
R=-\alpha \phi^2  F^2.
\end{equation}
In case of $\alpha=0$, one finds $R=0$ as in the $\alpha=0$ EMCS theory~\cite{Bekenstein:1974sf,Xanthopoulos:1991mx}. This is so  because one uses the conformally coupled scalar equation of $\nabla^2\phi-R\phi/6=0$ when taking the trace of Einstein equation, even though $T^{\phi}_{\mu\nu}$ is not traceless.

Making use of (\ref{ricciz}), one obtains the scalar equation
\begin{equation} \label{scalar-eq}
\nabla^2\phi+\frac{\alpha}{6}\Big[\phi^2-\frac{1}{\beta}\Big]  F^2 \phi=0.
\end{equation}
In case of $\alpha=0$, one finds  $\nabla^2\phi=0$ from Eq.(\ref{scalar-eq}). Also,  plugging  $R=0$ to Eq.(\ref{ascalar-eq}) with $\alpha=0$ leads to a  minimally coupled scalar equation ($\nabla^2\phi=0$)~\cite{Bekenstein:1974sf,Xanthopoulos:1991mx}. This  shows a feature of the $\alpha=0$ EMCS theory.

\section{$\alpha=0$ EMCS theory}
\subsection{Black hole solutions}

In this case, four equations of motion are obtained from the $\alpha=0$ EMCS theory
\begin{equation} \label{al0-eqs}
G_{\mu\nu}=2T^{\rm M}_{\mu\nu}+T^{\rm \phi}_{\mu\nu},~~\nabla^\mu F_{\mu\nu}=0,~~R=0,~~\nabla^2\phi=0.
\end{equation}
 We find  a constant scalar hairy (charged) black hole given by~\cite{Astorino:2013sfa}
\begin{eqnarray}
&&ds^2_{\rm csh}=\bar{g}_{\mu\nu}dx^\mu dx^\nu=-N(r)dt^2+\frac{dr^2}{N(r)}+r^2d\Omega^2_2, \nonumber \\
&&N(r)=1-\frac{2m}{r}+\frac{Q^2+Q^2_s}{r^2},~~\bar{\phi}_c=\pm \sqrt{\frac{1}{\beta}}\sqrt{\frac{Q^2_s}{Q^2_s+Q^2}},~~\bar{A}_t=\frac{Q}{r}-\frac{Q}{r_+}. \label{bbmb2}
\end{eqnarray}
which  are derived from solving the background equation of $\bar{G}_{\mu\nu}=2\bar{T}^{\rm M}_{\mu\nu}/(1-\beta \bar{\phi}_c^2)=2\bar{T}_{\mu\nu}$ together with $\bar{T}^{\mu}~_\nu=\frac{Q^2+Q^2_s}{r^4}{\rm diag}[-1,-1,1,1]$ and $\bar{\phi}_c={\rm const}$. From now on, the overbar ($~\bar{}~$) represents the background spacetime (solution to background equations).
Here $\beta=\kappa/6=4\pi G/3$.  The positions of outer and inner horizons
are given by $r_{\pm}=m\pm \sqrt{m^2-Q^2-Q^2_s}$.
This black hole has   a primary scalar hair and its geometry is similar to a non-extremal RN black hole except that the position of the horizon is shifted  by the presence of the scalar charge $Q_s$. However, we wish to point out that $\bar{\phi}_c$ depends on both $Q_s$ and $Q$, implying that it is not strictly a primary hair.
Its thermodynamics was  well established when replacing the Newtonian constant $G=3\beta/(4\pi)=1$ with the effective constant $\tilde{G}=G(Q^2+Q^2_s)/Q^2$:  ADM mass $M=\frac{m}{\tilde{G}}$  and entropy $S=\frac{A}{4\tilde{G}}$. The ADM mass and entropy go to zero when the charge $Q$ approaches zero.  This suggests that the constant scalar hairy black hole cannot radiate away its charge $Q$ and
settle down to a constant scalar hairy black hole (uncharged). In other words, one could not find a constant scalar black hole from the Einstein-conformally coupled scalar theory and thus, the BBMB solution  is the only scalar hair black hole~\cite{Bocharova:1970skc,Bekenstein:1974sf}.
The stability of the constant scalar hair  black hole will be explored in the next section.

Before we proceed, it is interesting to mention  the other solution named by the charged BBMB black hole for $\alpha=0$~\cite{Bekenstein:1974sf}
\begin{eqnarray}
&&ds^2_{\rm cBBMP}=-\Big(1-\frac{m}{ r}\Big)^2dt^2+\frac{dr^2}{\Big(1-\frac{m}{r}\Big)^2}+r^2d\Omega^2_2, \nonumber \\
&&\bar{\phi}(r)=\pm \sqrt{\frac{1}{\beta}}\frac{Q_s}{r-m},~~\bar{A}_t=\frac{Q}{r}-\frac{Q}{r_+},~~m=\sqrt{Q^2+Q_s^2}, \label{bbmb1}
\end{eqnarray}
where $m$ is the mass of the black hole. Here, the scalar hair $\bar{\phi}(r)$ is still the solution to $\bar{\nabla}^2\bar{\phi}=0$.  This line element takes  the same form as in the extremal RN black hole, but
the scalar hair blows up at the horizon $r=m$ and it belongs to the secondary hair.
It was shown   forty years ago that this black hole is unstable against the scalar perturbation~\cite{Bronnikov:1978mx} probably since it belongs to an extremal RN black hole~\cite{Onozawa:1995vu}.

\subsection{Stability of constant scalar hairy  black hole}
In the $\alpha=0$ EMCS theory, all perturbations around the background spacetime are introduced  as
\begin{equation} \label{lin-eqcs}
g_{\mu\nu}=\bar{g}_{\mu\nu}+h_{\mu\nu},~~A_\mu=\bar{A}_\mu+a_\mu,~~\phi=\bar{\phi}_c+\varphi.
\end{equation}
Substituting (\ref{lin-eqcs}) into  Eqs.(\ref{nequa1}), (\ref{maxwell-eq}), and (\ref{scalar-eq}) together  with $\alpha=0$,
their linearized equations around (\ref{bbmb2}) are given by
\begin{eqnarray}
&&\delta G_{\mu\nu}(h)=\delta T^{\rm \phi}_{\mu\nu}(h,\varphi)+2\delta T_{\mu\nu}^{\rm M}(h,f),  \label{csh-peq1} \\
&&\bar{\nabla}^{\mu}f_{\mu\nu}=0, \label{csh-peq2} \\
&&\bar{\nabla}^2\varphi  =0, \label{csh-peq2}
\end{eqnarray}
where
\begin{eqnarray}
&&\delta G_{\mu\nu}(h)=\delta R_{\mu\nu}(h)-\frac{1}{2} \delta R(h) \bar{g}_{\mu\nu}, \nonumber \\
&&\delta T^{\rm \phi}_{\mu\nu}(h,\varphi)=\beta \bar{\phi}^2_c[\delta G_{\mu\nu}(h)+2\bar{G}_{\mu\nu} \varphi-2 \bar{\nabla}_\mu\bar{\nabla}_\nu \varphi], \label{lin-seq} \\
&&\delta T_{\mu\nu}^{\rm M}(h,f)=2\bar{F}_{(\nu}~^\rho f_{\mu)\rho}-\bar{F}_{\mu\rho}\bar{F}_{\nu\sigma}h^{\rho\sigma}+\frac{1}{2}(\bar{F}_{\kappa\eta}f^{\kappa\eta}
-\bar{F}_{\kappa\eta}\bar{F}^{\kappa}~_{\sigma}h^{\eta \sigma})\bar{g}_{\mu\nu}-\frac{1}{4}\bar{F}^2 h_{\mu\nu},\nonumber \\
&&f_{\mu\nu}=\partial_\mu a_\nu-\partial_\nu a_\mu. \nonumber
\end{eqnarray}
We note that  Eq.(\ref{csh-peq1}) becomes a coupled tensor-vector-scalar equation.
Also, it is important to note that the scalar perturbation $\delta T^\phi_{\mu\nu}(h,\varphi)$  contributes to  the polar sector only.
Since the odd sector is the same as that for the EMS theory~\cite{Myung:2019oua}, we do not consider the odd sector here.
We may expand the metric  perturbation $h_{\mu\nu}$ in terms of tensor spherical  harmonics by choosing the Regge-Wheeler gauge.
For our purpose, we introduce four tensor modes ($H_0,~H_1,~H_2,~K$), two vector modes ($u_1,u_2$), and single scalar mode $\delta \phi_1$ in $ \varphi=\int d\omega e^{-i\omega t} \delta \phi_1(r) Y^l_m(\theta,\phi)$ for polar sector~\cite{Myung:2019oua}.
In this case, the polar sector of Eqs. (\ref{csh-peq1})-(\ref{csh-peq2}) is given by six coupled equations as
\begin{eqnarray}
K'(r)&=&-\left(\frac{l(l+1)+2N+2r N'-2}{2r^2}-\frac{\bar{A}_t'^2}{\bar{\phi}^2_c-1}\right)H_1(r)+\frac{H_0(r)}{r}
+\left(\frac{N'}{2N}-\frac{1}{r}\right)K(r)\nonumber\\
&&+\frac{\bar{\phi}_c[rN'\delta\phi_1(r)+2N(\delta\phi_1(r)-r\delta\phi_1'(r))]}{r^2N(\bar{\phi}^2_c-1)},\label{pol-eq1} \\
H_1'(r)&=&-\frac{4i \bar{A}_t'}{\omega(1-\bar{\phi}^2_c)}f_{12}(r)-\frac{H_0(r)+K(r)+N'H_1(r)}{N}
+\frac{4\bar{\phi}_c}{rN(1-\bar{\phi}^2_c)}\delta\phi_1(r), \label{pol-eq2}\\
H_0'(r)&=&\left(\frac{1}{r}-\frac{N'}{N}\right)H_0(r)-\left(\frac{1}{r}-\frac{N'}{2N}\right)K(r)
+\frac{4\bar{A}_t'}{(1-\bar{\phi}^2_c)N}f_{02}(r)+\frac{2\bar{\phi}_c\delta\phi_1'(r)}{r(1-\bar{\phi}^2_c)} \label{pol-eq3}\\
&&+\left(\frac{\omega^2}{N}-\frac{\bar{A}_t'^2}{1-\bar{\phi}^2_c}-\frac{l(l+1)}{2r^2}-\frac{N+rN'-1}{r^2}\right)H_1(r)
+\frac{\bar{\phi}_c(rN'-6N)}{r^2N(\bar{\phi}^2_c-1)}\delta\phi_1(r),\nonumber\\
f_{02}'(r)&=&\bar{A}_t'K(r)+\left(\frac{l(l+1)N}{r^2\omega}-\omega\right)if_{12}(r),\label{pol-eq4}\\
f_{12}'(r)&=&-\frac{i\omega}{N^2}f_{02}(r)-\frac{N'}{N}f_{12}(r),\label{pol-eq5}\\
\delta\phi''_1(r)&=&\Big[\frac{l(l+1)}{r^2N}-\frac{\omega^2}{N^2}
+\frac{N'}{rN}\Big]\delta\phi_1(r)-\frac{N'}{N}\delta\phi'_1(r),\label{pol-eq6}
\end{eqnarray}
where $H_2(r)=H_0(r)$ and $f_{01}(r)=i\omega f_{12}(r)+f'_{02}(r)$ with $f_{12}=u_2/rN$ and $f_{01}=u_1/r$.

The full analysis of stability could be performed by obtaining quasinormal frequency $\omega=\omega_r+i\omega_i$ for  physically propagating modes
when solving the above linearized equations with boundary conditions: ingoing modes at the outer horizon and purely outgoing modes at infinity.
If all $\omega_i$ are negative (no exponentially growing modes), the considering black hole is stable against all physically propagating modes.
Otherwise, the black hole is unstable.
We will compute the lowest quasinormal modes around the constant scalar hairy  black hole by making use of a direct-integration method.
Interestingly, the polar $l=2$ case includes three physically propagating modes: scalar-led, vector-led, and gravitational-led.
\begin{figure*}[t!]
   \centering
   \includegraphics{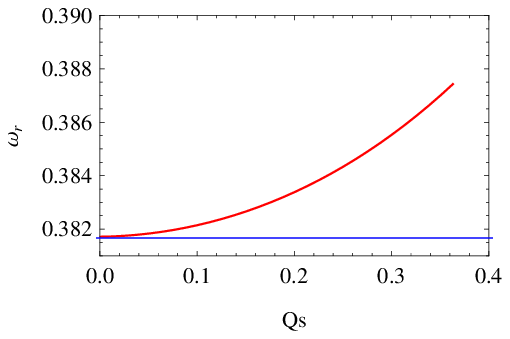}
      \hfill%
    \includegraphics{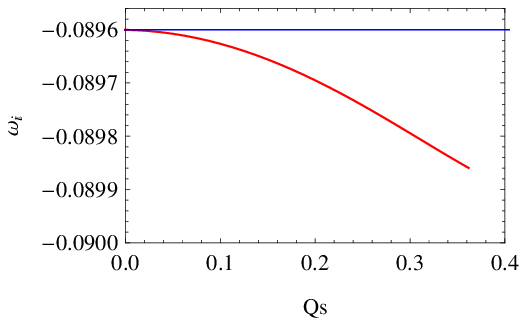}
\caption{The real (Left) and imaginary (Right) frequencies as function of scalar charge $Q_s$ for polar $l=2$ gravitational-led mode around the
constant scalar hairy black hole. The blue line of $0.38167 - i0.08961$ starting from $Q_s=0$ represents the quasinormal frequency for a polar $l=2$ gravitational-led mode propagating around the
RN  black hole.  }
\end{figure*}
From Fig. 1(Right), we observe that the imaginary  frequencies as function of scalar charge $Q_s$  are negative for polar $l=2$ gravitational-led mode, implying the stability
of the  constant scalar hairy  black hole.
Also, we have shown  that the  constant scalar hairy  black hole is stable against seven physically propagating modes of one for $l=0$ case, two for $l=1$ case, and four for $l=2$ case excluding  polar $l=2$ gravitational-led mode.

\section{$\alpha\not=0$ black hole solutions}

In this section, we wish to derive $\alpha\not=0$ black hole solutions numerically because any analytical solutions seem not to be allowed for the EMCS theory.
In 4.1, we will obtain the infinite scalarized black holes from the instability of RN black hole through the spontaneous scalarization.
In this case, the linearized scalar equation plays an important role of finding infinite bifurcation points for scalar clouds.
On the other hand, in 4.2, one could derive one branch of scalarized charged black holes by considering a constant scalar hairy black hole.
Finally, in 4.3, we  try to find another scalarized charged black holes by implementing the charged BBMB black hole. However, we fail to find a numerical solution for $\alpha\not=0$, whereas we obtain a numerical charged BBMB black hole solution for $\alpha=0$.

\subsection{ Infinite scalarized black holes from RN black hole}

For the EMCS theory, an allowed solution is  given by the RN black hole without scalar  hair
\begin{eqnarray}
&&ds^2_{\rm RN}=-f(r)dt^2+\frac{dr^2}{f(r)}+r^2d\Omega^2_2, \nonumber \\
&&f(r)=1-\frac{2m}{r}+\frac{Q^2}{r^2},~~\bar{\phi}=0,~~\bar{A}_t=\frac{Q}{r}-\frac{Q}{r_+},~~\bar{R}=0. \label{rnbh}
\end{eqnarray}
Here we would like to mention that  the condition of zero scalar ($\bar{\phi}=0$) is important to obtain the RN black hole solution.
The new scalarized charged black holes  may be  found from the appearance of instability for the RN black hole.

The linearized equations around a non-extremal RN black hole background  with $q=Q/m=0.7(Q=0.35,~m=1/2,~r_+=0.875$) are given by considering the  perturbations of tensor  $h_{\mu\nu}$, vector  $a_\mu$, and scalar $\varphi$ as
\begin{eqnarray}
\delta G_{\mu\nu}=2\delta T_{\mu\nu}^{\rm M},~~
\bar{\nabla}^{\mu}f_{\mu\nu}=0,~~
\Big[\bar{\nabla}^2 +\frac{\alpha}{3\beta}\frac{ Q^2}{r^4}\Big]\varphi=0, \label{RN-peq}
\end{eqnarray}
where the last (non-minimally coupled) scalar equation is important  to analyze the stability of the RN black hole because the first-two equations correspond to
 the Einstein-Maxwell linearized theory which are turned out to be stable against tensor-vector perturbations around the RN black hole background.

For the scalar perturbation, the separation of variables  is introduced around a spherically symmetric RN background (\ref{rnbh}) as
\begin{equation} \label{scalar-sp}
\varphi(t,r,\theta,\chi)=\frac{u(r)}{r}e^{-i\omega t}Y_{lm}(\theta,\chi).
\end{equation}
Considering a tortoise coordinate $r_*$ defined by $r_*=\int dr/f(r)$, a radial scalar equation is given by
\begin{equation} \label{sch-2}
\frac{d^2u}{dr_*^2}+\Big[\omega^2-V_{\rm u}(r)\Big]u(r)=0.
\end{equation}
Here the scalar potential $V_{\rm u}(r)$ is given by
\begin{equation} \label{pot-c}
V_{\rm u}(r)=\Big(1-\frac{2m}{r}+\frac{Q^2}{r^2}\Big)\Big[\frac{2m}{r^3}+\frac{l(l+1)}{r^2}-\frac{2Q^2}{r^4}-\frac{\alpha}{3\beta}\frac{Q^2}{r^4}\Big].
\end{equation}
The $s(l=0)$-mode is an allowable mode  for the scalar perturbation and thus,  it could be  used to test the instability of the RN black hole.
In this case, the sign of the last term is important to find the stability of the RN black hole. If $\beta<0$,
the potential is positive definite, leading to the stable black hole. If the sign is positive $(\beta>0$), it may induce a negative region outside the horizon, arriving at the unstable RN black hole. Hence, we choose $\beta=1/3$ to use the previous results for the EMS theory~\cite{Myung:2018vug}.
A dynamical scalar equation with $\varphi(t,r)=e^{\Omega t}\varphi(r)$ determines the  instability bound as $\alpha(q) >\alpha_{\rm th}(q)$ where
the threshold of instability is given by  $\alpha_{\rm th}(q=0.7)=8.019$. Actually, solving the static scalar equation with $\Omega=0$ leads to infinite  bifurcation points for scalar clouds
as $\alpha_n(q=0.7)=\{8.019,40.84,99.89,\cdots\}$ where $n$ indicates the appearance of $n=0,1,2,\cdots$ scalarized charged black holes.

In order to find scalarized charged black holes through the spontaneous scalarization, one assumes the metric and fields
\begin{eqnarray}
&&ds^2_{\rm scbh}=-\bar{N}(r)e^{-2\delta(r)}dt^2+\frac{dr^2}{\bar{N}(r)}+r^2d\Omega^2_2, \nonumber \\
&&\bar{N}(r)=1-\frac{2m(r)}{r},~~\bar{\phi}(r),~~\bar{A}_t=v(r). \label{sRN}
\end{eqnarray}
Substituting the above into Eqs.(\ref{nequa1}), (\ref{maxwell-eq}), and (\ref{scalar-eq}), one finds four equations for $m(r),\delta(r),v(r)$, and $\bar{\phi}(r)$:
\begin{eqnarray}
&&3e^{2\delta}r^2\alpha\bar{\phi}(\bar{\phi}^2-3)v'^2-18(m-m'r)\bar{\phi}
-r(r-2m)(9+\bar{\phi}^2)\bar{\phi}''\nonumber\\
&&-(r-2m)[\bar{\phi}(\bar{\phi}^2-3)\delta'
+(18+r(\bar{\phi}^2-9)\delta')\bar{\phi}'-2r\bar{\phi}\bar{\phi}'^2]=0,\label{neom1}\\
&&3e^{2\delta}r^2(1+\alpha\bar{\phi}^2)v'^2+2(r-2m)(\bar{\phi}^2-3)\delta'
+2\bar{\phi}[3m-2r+r(r-2m)\delta']\bar{\phi}'\nonumber\\
&&-3r(r-2m)\bar{\phi}'^2+2m'(\bar{\phi}^2+r\bar{\phi}\bar{\phi}'-3)=0,\label{neom2}\\
&&\Big(2+r\delta'+\frac{2r\alpha\bar{\phi}\bar{\phi}'}{1+\alpha\bar{\phi}^2}\Big)v'+rv''=0,\label{neom3}\\
&&-2r\bar{\phi}'^2+\delta'(\bar{\phi}^2-3+r\bar{\phi}\bar{\phi}')+r\bar{\phi}\bar{\phi}''=0, \label{neom4}
\end{eqnarray}
where the prime ($'$) denotes differentiation with respect to $r$.
The electric potential $v(r)$ can be eliminated from the above equations  by noticing  a first integral of
$v'(r)=\frac{Q e^{-\delta}}{r^2(1+\alpha\bar{\phi}^2)}$ with $Q$ electric charge. Also, this   implies  a vanishing electric potential in the near-horizon [$v_0=0$ in Eq.(\ref{aps-4})].

Accepting  an outer horizon located at $r=r_+$,  one may  introduce an
approximate solution to (\ref{neom1})-(\ref{neom4}) in the near-horizon region
\begin{eqnarray}\label{nexpr}
&&m(r)=\frac{r_+}{2}+m_1(r-r_+)+\cdots,\label{aps-1}\\
&&\delta(r)=\delta_0+\delta_1(r-r_+)+\cdots,\label{aps-2}\\
&&\bar{\phi}(r)=\phi_0+\phi_1(r-r_+)+\cdots,\label{aps-3}\\
&&v(r)=v_1(r-r_+)+\ldots,\label{aps-4}
\end{eqnarray}
where the coefficients are determined  by
\begin{eqnarray}
&&m_1=\frac{[(\alpha\phi_0^2(\phi_0^2-12)-9]Q^2}{6r_+^2(\phi_0^2-3)(1+\alpha\phi_0^2)^2},\quad
v_1=-\frac{e^{-\delta_0}Q}{r_+^2(1+\alpha\phi_0^2)},\nonumber\\
&&\phi_1=\frac{\alpha\phi_0 Q^2(\phi_0^2-3)^2}{r_+Q^2\left(9-\alpha\phi_0^2(\phi_0^2-12)\right)
+3r_+^3(\phi_0^2-3)(1+\alpha\phi_0^2)^2},\label{nceof}\\
&&\delta_1=\frac{\alpha\phi_0^2 Q^2(\phi_0^2-3)}{2r_+(1+\alpha\phi_0^2)\Big[Q^2(9-\alpha\phi_0^2(\phi_0^2-12))
+3r_+^2(\phi_0^2-3)(1+\alpha\phi_0^2)^2\Big]^2} \nonumber \\
&&\times \Big[12r_+^2(\phi_0^2-3)(1+\alpha\phi_0^2)^3+
Q^2\{18+\alpha(27+(48+63\alpha)\phi_0^2+(6\alpha-7)\phi_0^4
-3\alpha\phi_0^6)\}\Big].\nonumber
\end{eqnarray}
Here we note that $\delta_1$ takes the complicated form because of a conformal coupling term $\phi^2 R$.
This near-horizon solution involves  two essential  parameters of  $\phi_0=\bar{\phi}(r_+,\alpha)$ and $\delta_0=\delta(r_+,\alpha)$, which can be
determined  by  matching  (\ref{aps-1})-(\ref{aps-4}) with the asymptotic  solution in the far-region
\begin{eqnarray}\label{insol}
m(r)&=&M-\frac{3Q^2+Q_s^2}{6r}+\ldots, \quad \bar{\phi}(r)=\frac{Q_s}{r}+\cdots, \nonumber \\
\delta(r)&=&\frac{Q_s^2[2Q_s^2-6M^2+3Q^2(2+\alpha)]}{108r^4}+\cdots,\quad
v(r)=\Phi+\frac{Q}{r}+\cdots,
\end{eqnarray}
where the ADM mass $M$, the scalar charge $Q_s$, and the electrostatic potential $\Phi$ are included. Importantly, we mention that  any constant scalar is not allowed for the far-region.

We may determine the infinitely numerical solutions labeled by $n=0(\alpha\ge 8.019),~n=1(\alpha\ge 40.84),~n=2(\alpha \ge 99.89),\cdots$ scalarized charged black holes. Now,  let us  display $n=0,~1,~2$ branches for black holes in terms of scalar hair $\phi_0$ on the horizon in Fig. 2.
\begin{figure*}[t!]
   \centering
   \includegraphics{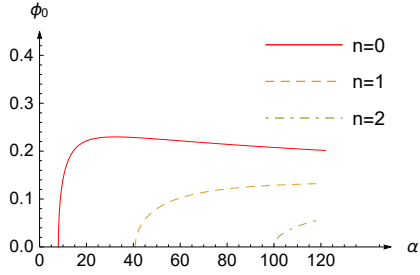}

\caption{The scalar hair $\phi_0(\alpha)=\bar{\phi}(r_+,\alpha)$ on the horizon as function of the coupling $\alpha$. Here, we display the first three branches among infinite branches. The $n=0$ branch starts from the first bifurcation point at $\alpha=8.019$,
whereas $n=1$ and 2 branches start from the second point at $\alpha=40.84$ and from the third point at $\alpha=99.89$. }
\end{figure*}

Explicitly, we choose the  horizon radius $r_+=0.857$ and electric charge $Q=0.35$
to  construct the $n=0$ scalarized charged black hole with $\alpha=65.25$ shown in Fig. 3.
An interesting behavior is found from $\delta(r)$, compared to that in the EMS theory.
It is noted that $\delta(r)$ in (\ref{insol}) takes a different form  from $\delta(r)=2Q_s^2/r^2$ in the EMS theory~\cite{Myung:2018vug}.
\begin{figure*}[t!]
   \centering
   \includegraphics{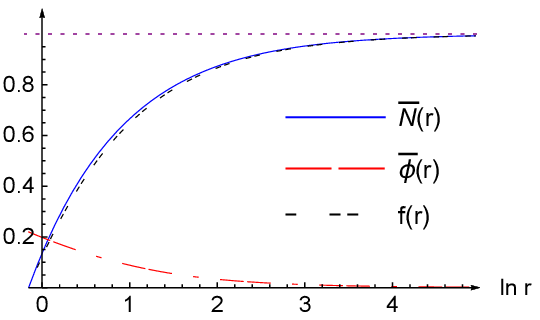}
      \hfill%
    \includegraphics{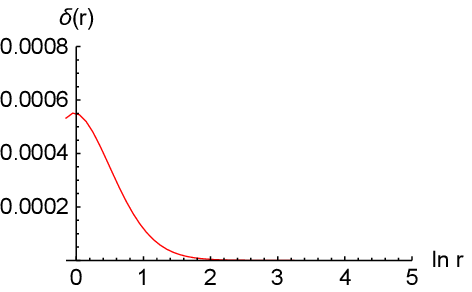}
\caption{Plot of a scalarized charged black hole with $\alpha=65.25$ residing in the  $n=0$ branch. Here $f(r)$ represents the metric function for the RN black hole.
We note $\delta(r)=0$ for the RN black hole.
 We plot all in terms of  `$\ln r$' and thus,
 the horizon is located at $\ln r=\ln r_+=-0.154$.}
\end{figure*}

\subsection{Scalarized black hole  inspired by constant scalar hairy black hole}
Let us derive the scalarized charged black hole inspired by  the  constant scalar hairy black hole.
In this case, even though the near-horizon expansion is  the same as  (\ref{aps-1})-(\ref{aps-4}),
we find a different far-region expansion, compared with (\ref{insol}) as
\begin{eqnarray}\label{insolC}
m(r)&=&M-\frac{3Q_s^2(1+\phi_\infty^2)}{2(\phi_\infty^2-3)^2r}
-\frac{Q^2(2\alpha\phi_\infty^4-15\alpha\phi_\infty^2-9)}{6(\phi_\infty^2-3)(1+\alpha\phi_\infty^2)^2r}
-\frac{MQ_s\phi_\infty}{(\phi_\infty^2-3)r}\cdots, \nonumber \\
\delta(r)&=&\frac{2Q_s\phi_\infty}{(\phi_\infty^2-3)r}+\ldots,\quad
v(r)=\Phi+\frac{Q}{(1+\alpha\phi_\infty^2)r}+\cdots, \nonumber \\
 \bar{\phi}(r)&=&\phi_\infty+\frac{Q_s}{r}+\cdots.
\end{eqnarray}
In case of $\phi_\infty=0$, the above reduces to (\ref{insol}).
Here, we may choose  $\phi^2_\infty$  to be $\bar{\phi}^2_c=\frac{3Q_s^2}{Q^2+Q_s^2}<3$ as in the constant scalar hairy black hole (\ref{bbmb2}).
However, it is important to note  that the series  solution (\ref{insolC}) is quite different from the analytic solution (\ref{bbmb2}).
For the constant scalar hairy black hole,
we may  select electric charge $Q=0.2$, scalar charge $Q_s=0.1$, and mass $M=1/2$, providing  the horizon radius $r_+=0.9472$ and $\bar{\phi}_c= 0.7746$.
Now, we also choose the same horizon radius $r_+=0.9472(\ln r_+=-0.0542)$ and electric charge $Q=0.2$ to construct a single branch of  scalarized charged black holes existing from $\alpha=0$ ($\phi_0=\phi_\infty$) to  $\alpha=\infty$ ($\phi_0=\phi_\infty=0.7746$) in Fig. 4. This single branch is compared to many branches in Fig. 2.
\begin{figure*}[t!]
   \centering
   \includegraphics{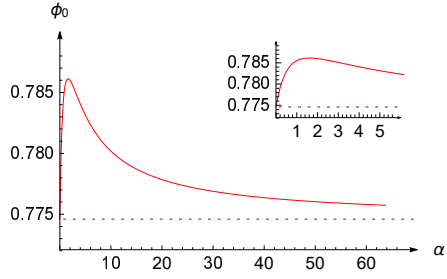}

\caption{Plot of  scalar hair $\phi_0(\alpha)=\bar{\phi}(r_+,\alpha)$ on the horizon as function of the coupling $\alpha\in[0,\infty)$, showing a single branch.
The dashed line denotes $\phi_\infty=\bar{\phi}_c=0.7746$.
The magnification indicates that $\phi_0$ starts with $\phi_\infty$. }
\end{figure*}

Explicitly, we wish to display a scalarized charged black hole with $\alpha=63.75$ in Fig.  5.
$\bar{N}(r)$ and $N(r)$ represent metric function for scalarized charged black hole and constant scalar hairy black hole, respectively. The magnification in the left picture shows an enlarged tendency of scalar hair $\bar{\phi}(r)$ clearly, differing from a constant hair $\bar{\phi}_c=0.7746$ in the constant scalar hairy black hole (\ref{bbmb2}). From the right picture of Fig. 5, we observe that $\delta(r)$ is negative, while $\delta(r)=0$ for the constant scalar hairy black hole.

\begin{figure*}[t!]
   \centering
   \includegraphics{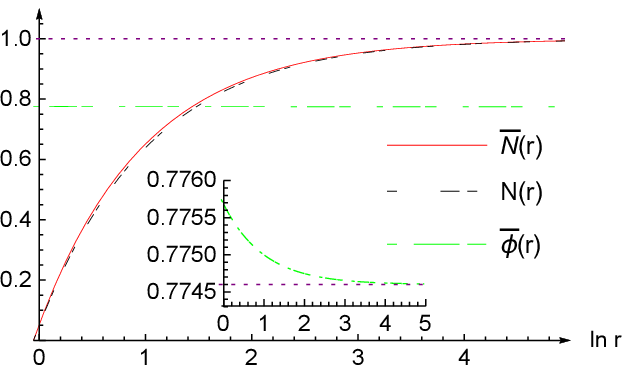}
         \hfill%
    \includegraphics{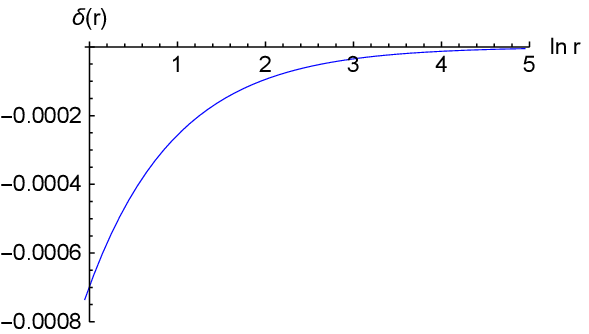}
\caption{ Plot of a scalarized black hole with  $\alpha=63.75$, compared to the constant scalar hairy black hole [$N(r),~\bar{\phi}_c=0.7746,~\delta(r)=0$]. The horizon is located at $\ln r=\ln r_+=-0.0542$.}
\end{figure*}

\subsection{Numerical charged BBMB black hole}
We try to find another scalarized charged black holes  by  considering  the charged BBMB solution (\ref{bbmb1}).
Taking into account metric and fields in (\ref{sRN}), we consider the expansion in the near-horizon region as
\begin{eqnarray}\label{CBBMBexpr2}
&&\bar{N}(r)=\sum^{\infty}_{i=2} N_i(r-r_+)^i,\quad
\delta(r)=\sum^{\infty}_{i=0}\delta_i(r-r_+)^i,\nonumber\\
&&\bar{\psi}(r)=\sum^{\infty}_{i=1}\psi_i(r-r_+)^i,\quad
v(r)=\sum^{\infty}_{i=1}v_i(r-r_+)^i,
\end{eqnarray}
where we introduce $\bar{\psi}(r)=1/\bar{\phi}(r)$ for a technical computation.
Replacing  $\bar{\phi}(r)$ by $1/\bar{\psi}(r)$, all equations (\ref{neom1})-(\ref{neom4}) take the forms
\begin{eqnarray}
&& \alpha r e^{2\delta}(1/3-\bar{\psi}^2)v'^2+\bar{\psi}'[r\bar{\psi} \bar{N}'-\bar{N}(\bar{\psi}(r\delta'-2)+2r\bar{\psi}')]+r\bar{N}+\bar{\psi}\bar{\psi}''=0,\label{maxN1}\\
&&(1-3\bar{\psi}^2)[1+\bar{N}(2r\delta'-1)]-\frac{r\bar{N}\bar{\psi}'}{\bar{\psi}^2}[2\bar{\psi}(r\delta'-2)+3r\bar{\psi}']\nonumber\\
&&+3r^2 e^{2\delta}(\alpha+\bar{\psi}^2)v'^2+\frac{r\bar{N}'}{\bar{\psi}}(r\bar{\psi}'+3\bar{\psi}^3-\bar{\psi})=0,\label{maxN2}\\
&&Q\bar{\psi}^2+e^{\delta}r^2(\alpha+\bar{\psi}^2)v'=0,\label{maxN3}\\
&&\delta'(3\bar{\psi}^3-\bar{\psi}+r\bar{\psi}')+r\bar{\psi}''=0. \label{maxN4}
\end{eqnarray}
Substituting (\ref{CBBMBexpr2}) into (\ref{maxN1})-(\ref{maxN4}) leads to the lowest order equations
\begin{equation}
1-r_+^2N_2+3\alpha e^{2\delta_0}r_+^2 v_1^2=0,\quad \alpha e^{2\delta_0}r_+ v_1^2=0,\quad \alpha e^{\delta_0}r_+^2 v_1=0,\quad \delta_1 \psi_1+2\psi_2=0.
\end{equation}
However,  considering $r_+\not=0$ and $v_1\not=0$, the second and third equations imply
\begin{equation}
 \alpha=0.
 \end{equation}
 This means that the EMCS theory is not compatible  with expansion (\ref{CBBMBexpr2}) when deriving a scalarized extremal  black hole like the charged BBMB solution.
 We need to look for  another expansion  which is suitable for constructing a scalarized extremal black hole solution.

In this section, instead, we wish to derive the charged BBMB black hole solution (\ref{bbmb1}) numerically by considering the $\alpha=0$ case
because it is important to know  what  the exact nature of the charged BBMB solution is.
The first three coefficients in the near-horizon region are determined  by
\begin{eqnarray}\label{ncoeff2}
&& N_2=\frac{1}{r_+^2},\quad N_3=-\frac{6(r_+^2-Q^2)\psi_1^2+4}{3r_+^3},\nonumber\\
&&N_4=\frac{5+15(r_+^2-Q^2)\psi_1^2}{3r_+^4}-\frac{3(r_+^2-Q^2)^2\psi_1^4}{r_+^4},\nonumber\\
&&\delta_0,\quad \delta_1=\frac{2[1-3(r_+^2-Q^2)\psi_1^2]}{3r_+},\quad
\delta_2=-\frac{1-15(r_+^2-Q^2)\psi_1^2+36(r_+^2-Q^2)^2\psi_1^4}{9r_+^2},\nonumber\\
&&\psi_1, \quad \psi_2=-\frac{[1-3(r_+^2-Q^2)\psi_1^2]\psi_1}{3r_+},\quad \psi_3=\frac{2\psi_1[1-3(r_+^2-Q^2)\psi_1^2]^2}{9r_+^2}, \nonumber \\
&&v_1=-\frac{e^{-\delta_0}Q}{r_+^2},\quad
v_2=\frac{e^{-\delta_0}Q}{r_+^3}\Big[\frac{4}{3}-(r_+^2-Q^2)\psi_1^2\Big],\nonumber\\
&&v_3=-\frac{e^{-\delta_0}Q}{r_+^4}\Big[\frac{14}{9}-\frac{7}{3}(r_+^2-Q^2)\psi_1^2+2(r_+^2-Q^2)^2\psi_1^4\Big],
\end{eqnarray}
which are not  appropriate near-horizon forms for the charged BBMB black hole.
In this approach, two essential parameters $\delta_0$ and $\psi_1$ will be determined
when matching (\ref{ncoeff2}) with the following asymptotic solution
for $r\gg r_+$:
\begin{eqnarray}\label{insol2}
\bar{N}(r)&=&1-\frac{2m}{r}+\frac{Q^2+Q_s^2}{r^2}+\cdots, \quad
\bar{\phi}(r)=\frac{\sqrt{3}Q_s}{r}+\frac{\sqrt{3}Q_s m}{r^2}+\cdots, \nonumber \\
\delta(r)&=&\frac{Q_s^2(Q^2+Q_s^2-m^2)}{6r^4}+\cdots,\quad
v(r)=\frac{Q}{r}-\frac{Q}{r_+},
\end{eqnarray}
which are not surely  appropriate asymptotic forms for the charged BBMB black hole.
In deriving a numerical charged BBMB black hole solution, an important requirement is to implement  the asymptotic flatness.
We may choose $\delta_0=0$ when considering the asymptotic boundary at $r=\infty$.
Actually, we choose here $\delta_0=0.0001$ because the asymptotic boundary is located at $r=100$ for numerical computation.
We make use of  Table 1 which includes  numerical relations between ($r_+,~\psi_1,~Q_s,~m,~Q$)  for different $r_+$ with the same $Q$.
\begin{table}[h]
   \centering
\begin{tabular}{|c|c|c|c|c|}\hline
$r_+$  &$\psi_1$ &  $Q_s$ & $m$ & $Q$ \\ \hline
0.75&1.0327 & 0.5590&0.7501&0.5\\ \hline
1 & 0.6666 & 0.8860& 0.9999&0.5\\ \hline
1.25& 0.5039 & 1.1456& 1.2500 &0.5 \\ \hline
1.5& 0.4082 & 1.4142& 1.5000  &0.5\\ \hline
\end{tabular}
\caption{Table showing numerical relations  between ($r_+,~\psi_1,~Q_s,~m,~Q$)  for different $r_+$ with the same $Q=0.5$.
These include three relations: $r_+\approx m$, $m^2\approx Q^2+Q^2_s$, and $\psi_1\approx 1/\sqrt{3}Q_s$. }
\end{table}
Confirming three  relations from Table 1 as
\begin{eqnarray}
r_+\approx m,~m\approx \sqrt{Q^2+Q_s^2},~ \psi_1\approx\frac{1}{\sqrt{3(r_+^2-Q^2)}}\approx\frac{1}{\sqrt{3}Q_s},
\end{eqnarray}
we read off correct coefficients
\begin{eqnarray}
&&N_2=\frac{1}{r_+^2},\quad N_3=-\frac{2}{r_+^3},\quad N_4=\frac{3}{r_+^4},\cdots,\nonumber\\
&&\delta_i=0, ~{\rm for}~i=1,\cdots,\nonumber\\
&& \psi_i=0, ~{\rm for}~i=2,\cdots,\nonumber\\
&& v_1=-\frac{Q}{r_+^2},\quad v_2=\frac{Q}{r_+^3},\quad v_3=-\frac{Q}{r_+^4},\cdots.
\end{eqnarray}
Making use of the above relations, we arrive at  the near-horizon expansion  for the charged BBMB solution (\ref{bbmb1})
\begin{eqnarray}\label{ncoef2}
&&\bar{N}(r)=\frac{(r-r_+)^2}{r_+^2}-\frac{2(r-r_+)^3}{r_+^3}+\frac{3(r-r_+)^4}{r_+^4}+... \to \Big[\Big(1-\frac{r_+}{r}\Big)^2\Big]_{r=r_+},\nonumber\\
&&v(r)=-\frac{Q(r-r_+)}{r_+^2}+\frac{Q(r-r_+)^2}{r_+^3}-\frac{Q(r-r_+)^3}{r_+^4}+...\to \Big[\frac{Q}{r}\Big]_{r=r_+}-\frac{Q}{r_+}, \nonumber\\
&&\bar{\psi}(r)=\frac{r-r_+}{\sqrt{3(r_+^2-Q^2)}},\nonumber\\
&&  \delta(r)\approx 0.
\end{eqnarray}
On the other hand, using a relation of  $m^2\approx Q^2+Q^2_s$ leads to   the asymptotic  expansion for the  charged BBMB solution 
\begin{eqnarray}\label{insol2}
&&\bar{N}(r)=1-\frac{2m}{r}+\frac{m^2}{r^2}=\Big(1-\frac{m}{r}\Big)^2, \nonumber \\
&&\bar{\phi}(r)=\frac{\sqrt{3(r_+^2-Q^2)}}{r}+\frac{\sqrt{3(r_+^2-Q^2)}~ m}{r^2}+\cdots \to \Big[\frac{\sqrt{3(r_+^2-Q^2)}}{r-m}\Big]_{r\gg m}, \nonumber \\
&&v(r)=\frac{Q}{r}-\frac{Q}{r_+}, \nonumber \\
&&  \delta(r)\approx 0.
\end{eqnarray}
Eqs.(\ref{ncoef2}) and (\ref{insol2})  show clearly how the charged BBMB black hole could be obtained   from the $\alpha=0$ EMCS theory when imposing the asymptotically flat condition. Also, we note that these do not contain the scalar charge $Q_s$, implying that the scalar hair is secondary in the charged BBMB solution.
As far as we know, this is the first time to recover the charged BBMB solution from the series solution.

\section{Discussions}
We have  investigated  the EMCS theory to obtain various scalarized charged black hole solutions.
It is well known that the $\alpha=0$ EMCS theory has yielded   the constant scalar hairy  black hole and charged BBMB  black hole.
We has shown that  the former is stable against full perturbations. It is interesting to note  that  the latter remains unstable because it belongs to an extremal black hole~\cite{Bronnikov:1978mx}.

For $\alpha\not=0$, it is reasonable to say  that the unstable RN black holes  imply the appearance of $n=0(\alpha \ge 8.019),1(\alpha \ge 40.84),2(\alpha \ge 99.89),\cdots$ scalarized charged black holes. We could obtain infinite branches of scalarized charged black holes through the spontaneous scalarization.
Importantly, for $\alpha>0$, we obtain  a single branch of scalarized charged black hole  inspired by  the constant scalar hairy black hole.
Unfortunately, for $\alpha>0$, we have failed to derive another scalarized black hole by considering the charged BBMB black hole.

However, we have shown explicitly  how   the charged BBMB black hole could be found from the $\alpha=0$ EMCS theory when requiring  the condition of  asymptotic flatness.
This implies that the charged BBMB solution dictates the feature of conformally coupled scalar system in the $\alpha=0$ EMCS theory.

On the other hand, the EMS theory has provided infinite branches of scalarized charged black holes based on the instability  of RN black holes~\cite{Herdeiro:2018wub,Myung:2018vug}.

 \vspace{2cm}

{\bf Acknowledgments}
 \vspace{1cm}

This work was supported by the National Research Foundation of Korea (NRF) grant funded by the Korea government (MOE)
 (No. NRF-2017R1A2B4002057).


\newpage

\begin{thebibliography}{99}
\bibitem{Ruffini:1971bza}
  R.~Ruffini and J.~A.~Wheeler,
  Phys.\ Today {\bf 24}, no. 1, 30 (1971).
  doi:10.1063/1.3022513

\bibitem{Herdeiro:2015waa}
  C.~A.~R.~Herdeiro and E.~Radu,
  Int.\ J.\ Mod.\ Phys.\ D {\bf 24}, no. 09, 1542014 (2015)
  doi:10.1142/S0218271815420146
  [arXiv:1504.08209 [gr-qc]].

\bibitem{Bocharova:1970skc}
  N.~M.~Bocharova, K.~A.~Bronnikov and V.~N.~Melnikov,
  Vestn.\ Mosk.\ Univ.\ Ser.\ III Fiz.\ Astron.\ , no. 6, 706 (1970).

\bibitem{Bekenstein:1974sf}
  J.~D.~Bekenstein,
  Annals Phys.\  {\bf 82}, 535 (1974).
  doi:10.1016/0003-4916(74)90124-9

\bibitem{Xanthopoulos:1991mx}
  B.~C.~Xanthopoulos and T.~Zannias,
  J.\ Math.\ Phys.\  {\bf 32}, 1875 (1991).




\bibitem{Myung:2019adj}
  Y.~S.~Myung and D.~C.~Zou,
  Phys.\ Rev.\ D {\bf 100}, no. 6, 064057 (2019)
  doi:10.1103/PhysRevD.100.064057
  [arXiv:1907.09676 [gr-qc]].

\bibitem{Doneva:2017bvd}
  D.~D.~Doneva and S.~S.~Yazadjiev,
  Phys.\ Rev.\ Lett.\  {\bf 120}, no. 13, 131103 (2018)
  doi:10.1103/PhysRevLett.120.131103
  [arXiv:1711.01187 [gr-qc]].

\bibitem{Silva:2017uqg}
  H.~O.~Silva, J.~Sakstein, L.~Gualtieri, T.~P.~Sotiriou and E.~Berti,
  Phys.\ Rev.\ Lett.\  {\bf 120}, no. 13, 131104 (2018)
  doi:10.1103/PhysRevLett.120.131104
  [arXiv:1711.02080 [gr-qc]].

\bibitem{Antoniou:2017acq}
  G.~Antoniou, A.~Bakopoulos and P.~Kanti,
  Phys.\ Rev.\ Lett.\  {\bf 120}, no. 13, 131102 (2018)
  doi:10.1103/PhysRevLett.120.131102
  [arXiv:1711.03390 [hep-th]].



\bibitem{Herdeiro:2018wub}
  C.~A.~R.~Herdeiro, E.~Radu, N.~Sanchis-Gual and J.~A.~Font,
  Phys.\ Rev.\ Lett.\  {\bf 121}, no. 10, 101102 (2018)
  doi:10.1103/PhysRevLett.121.101102
  [arXiv:1806.05190 [gr-qc]].

\bibitem{Bronnikov:1978mx}
  K.~A.~Bronnikov and Y.~N.~Kireev,
  Phys.\ Lett.\ A {\bf 67}, 95 (1978).
  doi:10.1016/0375-9601(78)90030-0

\bibitem{Myung:2018vug}
  Y.~S.~Myung and D.~C.~Zou,
  Eur.\ Phys.\ J.\ C {\bf 79}, no. 3, 273 (2019)
  doi:10.1140/epjc/s10052-019-6792-6
  [arXiv:1808.02609 [gr-qc]].

\bibitem{Astorino:2013xc}
  M.~Astorino,
  Phys.\ Rev.\ D {\bf 87}, no. 8, 084029 (2013)
  doi:10.1103/PhysRevD.87.084029
  [arXiv:1301.6794 [gr-qc]].

\bibitem{Astorino:2013sfa}
  M.~Astorino,
  Phys.\ Rev.\ D {\bf 88}, no. 10, 104027 (2013)
  doi:10.1103/PhysRevD.88.104027
  [arXiv:1307.4021 [gr-qc]].

\bibitem{Onozawa:1995vu}
  H.~Onozawa, T.~Mishima, T.~Okamura and H.~Ishihara,
  Phys.\ Rev.\ D {\bf 53}, 7033 (1996)
  doi:10.1103/PhysRevD.53.7033
  [gr-qc/9603021].

\bibitem{Myung:2019oua}
  Y.~S.~Myung and D.~C.~Zou,
  Eur.\ Phys.\ J.\ C {\bf 79}, no. 8, 641 (2019)
  doi:10.1140/epjc/s10052-019-7176-7
  [arXiv:1904.09864 [gr-qc]].


\end{thebibliography}
\end{document}